\documentclass[conference]{IEEEtran}
\IEEEoverridecommandlockouts
\usepackage{cite}
\usepackage{amsmath,amssymb,amsfonts}
\usepackage{algorithmic}
\usepackage{graphicx}
\usepackage{textcomp}
\usepackage{xcolor}
\usepackage{booktabs}
\usepackage{url}
\usepackage{hyperref} 
\usepackage{xcolor}
\hypersetup{
    colorlinks=true, 
    linkcolor=blue, 
    filecolor=black, 
    urlcolor=blue, 
    citecolor=blue, 
    linkbordercolor={1 1 1}, 
    pdfborder={0 0 0} 
}
\def\BibTeX{{\rm B\kern-.05em{\sc i\kern-.025em b}\kern-.08em
T\kern-.1667em\lower.7ex\hbox{E}\kern-.125emX}}
\begin{document}

\title{Comparative Withholding Behavior Analysis of Historical Energy Storage Bids in California
\thanks{This work was partly supported by the National Science Foundation under award ECCS-2239046.}
}

\author{\IEEEauthorblockN{Neal Ma, Ningkun Zheng, Ning Qi, Bolun Xu}
\IEEEauthorblockA{\textit{Earth and Environmental Engineering} \\
\textit{Columbia University}\\
New York, USA\\
\{nam2252, nz2343, nq2176, bx2177\}@columbia.edu}
}

\maketitle

\begin{abstract}
The rapid growth of battery energy storage in wholesale electricity markets calls for a deeper understanding of storage operators' bidding strategies and their market impacts. This study examines energy storage bidding data from the California Independent System Operator (CAISO) between July 1, 2023, and October 1, 2024, with a primary focus on economic withholding strategies. Our analysis reveals that storage bids are closely aligned with day-ahead and real-time market clearing prices, with notable bid inflation during price spikes. Statistical tests demonstrate a strong correlation between price spikes and capacity withholding, indicating that operators can anticipate price surges and use market volatility to increase profitability. Comparisons with optimal hindsight bids further reveal a clear daily periodic bidding pattern, highlighting extensive economic withholding. These results underscore potential market inefficiencies and highlight the need for refined regulatory measures to address economic withholding as storage capacity in the market continues to grow.
\end{abstract}


\section{Introduction}

With the rapid integration of energy storage resources into power systems, especially within the California Independent System Operator (CAISO) and the Electric Reliability Council of Texas (ERCOT), understanding the strategic behavior of energy storage systems (ESS) and their impact on market operations is increasingly critical. From 2020 to 2024, CAISO interconnected over 10,000 MW of battery capacity \cite{CAISO_ES_2023}. With this growing influx of Battery Energy Storage Systems (BESS) in wholesale markets, most BESS units now prioritize price arbitrage within the energy market, charging during low-price periods, driven by zero-marginal cost solar and wind generation, and discharging during high-price periods, such as the rapid ramping intervals in the evenings in California.

BESS bids differ fundamentally from those of conventional generators due to the distinct physical and economic properties of storage systems \cite{zheng2023energy}. Unlike traditional bids based on fuel costs and system constraints, which are well-understood by system operators, BESS bids rely on private market condition forecasts and strategic timing to maximize profit potential \cite{zhou2024energy}. Market design studies \cite{harvey2001market} and CAISO regulations \cite{caiso_phase4_2023} acknowledge that energy storage bids should be guided by predicted price trends, given their limited energy supply and unique ability to target high-price periods, aligning with market conditions where supply-demand imbalances intensify. In CAISO, BESS operators are required to bid their full capacity, precluding physical withholding but allowing for economic withholding—bids crafted not on operational costs but on anticipated market opportunities.

While economic withholding by storage systems is legitimate within regulatory frameworks, it does not eliminate the risk of inefficiencies and potential market power abuse. Some BESS units may lack advanced forecasting tools or sophisticated bidding software, relying instead on ad hoc strategies. Meanwhile, others may exercise market power similarly to conventional generators, potentially inflating prices, making it challenging to differentiate between various bidding motivations. Although emerging research examines the strategic bidding and potential market power of storage systems \cite{sioshansi_when_2014, williams_electricity_2022, mohsenian-rad_coordinated_2016, ye_investigating_2019, chabok_assessment_2020, bhattacharjee_energy_2022}, little has been done to analyze the actual bidding behavior of current market-participating energy storage systems.

In this study, we collect public BESS bid data from CAISO and analyze bidding patterns with the following key findings:
\begin{itemize} 
\item BESS participants engage in extensive economic withholding to strategically charge during mid-day and discharge in late evenings, with withholding more prevalent in the day-ahead markets due to lower price volatility. 
\item Through comparisons with hindsight optimal bids generated using historical prices, we show that BESS bids are significantly inflated beyond what is required to reflect future opportunity value. 
\item This inflated withholding may reduce market efficiency, as BESS units can miss price spikes when these occur outside typical late evening discharge windows. We support our analysis with an extreme example day of Aug 16, 2024, in which storage failed to mitigate system price spikes that occurred outside their preferred discharge hours. 
\item Comparative analysis between price spike and non-spike days shows that BESS participants design bids in anticipation of price spikes, with notably higher discharge bids on spike days than on regular days. This raises concerns about market power exercise and underscores both the urgency and complexity of regulating storage market power. 
\end{itemize}

The rest of the paper is organized as follows. Section~\ref{method} introduces the methodology of data collection and analytic methods. Section~\ref{results} presents the analysis results, and Section~\ref{conclusion} discusses key findings and provides the conclusion.

\vspace{0.1cm}

\section{Methodology}\label{method}

\subsection{Data Collection}

We collect from published energy storage bid data from between July 1, 2023 until October 1, 2024. These data are scraped directly from \textit{Daily Energy Storage Reports}~\cite{caiso_daily_storage_reports} and include aggregated energy storage bids\footnote{Code and data: [https://github.com/nmadev/CAISO-EnergyStorage]}. Bid data are binned into 11 discrete segments, as detailed in Table~\ref{BidSegmentDef}, and a self-schedule segment, where BESS schedules available capacity and operates as price-takers. The self-schedule segment never exceeds $6$\% of the total bid capacity and thus is excluded from these analyses. Additionally, during the period of interest, there are no bids that fall in the highest-valued bid segment, segment $11$. While we primarily analyze charge and discharge bid segment data for both the integrated forward market (IFM) and real-time pre-dispatch (RTPD), the published data also includes energy and state of charge awards, ancillary service awards, and hybrid resource data. These data are aggregated across CAISO so we do not have transparency into the spatial distribution of bids, awards, and capacity.
\begin{table}[htbp]
    \caption{Bid Segment Definitions}
    \label{BidSegmentDef}
    \centering
    \begin{tabular}{c c c c}
    \toprule
    Segment Label & Range (\$/MW) & Segment Label & Range (\$/MW) \\
    \midrule
    1 & [-150, -100] & 7 & (50, 100] \\
    2 & (-100, -50] & 8 & (100, 200] \\
    3 & (-50, -15] & 9 & (200, 500] \\
    4 & (-15, 0] & 10 & (500, 1000] \\
    5 & (0, 15] & 11 & (1000, 2000] \\
    6 & (15, 50] & & \\
    \bottomrule
    \end{tabular}
\end{table}

From these bid segments, we define a weighted average bid from the average value of each individual bid segment, $v_i$, and proportion of total segmented bids, $p_i$, that a certain bid segment constitutes. The weighted average bid, $\overline{v}$, can be expressed as:
\begin{equation}
    \overline{v} = \sum_{i=1}^{11}p_iv_i, p_i = \frac{\text{Bid segment quantity in MW}}{\text{Total bid quantity in MW}} 
\end{equation}
This weighted average is sensitive to economic withholding for both charge and discharge bids. Discharge bids that are shifted to higher segments to withhold capacity will artificially inflate the average discharge bid while charge bids moved to negatively-priced segments to prevent charge capacity will lower the average bid price. We also define the bid spread at a given time as the difference between weighted average discharge and weighted average charge bids, or:
\begin{equation}
    \overline{v}_{\mathrm{spread}} = \overline{v}_{\mathrm{discharge}}-\overline{v}_{\mathrm{charge}}
\end{equation}
For periods where BESS operators are withholding capacity, we expect the bid spread to increase through the inflation of discharge bids, deflation of charge bids, or both.

Real-time and day-ahead price data are obtained for all three zones and four hubs in CAISO, respectively. Real-time price data are taken at 5-minute granularity while day-ahead prices are hourly. For each, a minimum, maximum, and average aggregation is constructed from the instantaneous minimum, maximum, and unweighted average price across all zones/hubs. Temporal aggregations are taken as the mean price over the aggregated time-period.

\subsection{Hindsight Optimal Bid Generation}

To evaluate the efficiency and strategy of historical storage bids, we generate hindsight optimal bids as a benchmark. We generate hindsight storage bids using an analytical dynamic programming method~\cite{zheng2022arbitraging}, based on historical CAISO price data from the same period. As the price influence of storage bids is hard to include in optimal bid analysis, we create hindsight storage bids that reflect optimal bidding strategy under historical market price and price taker assumption.

We formulate the energy storage arbitrage problem using a dynamic programming approach:
\begin{subequations}
\begin{align}
    Q_{t-1}(e_{t-1}) &= \max_{\substack{b_t\text{,} p_t\text{,} e_t \\ \in \mathcal{E}(e_{t-1})}} \lambda_t (p_t-b_t) - cp_t + Q_{t}(e_{t})
\end{align}
where the storage opportunity value is defined recursively as the maximized arbitrage profit, capturing the current time step's profit and future opportunity values. Here, $\lambda_t$ represents the price, $p_t$ and $b_t$ denote discharge and charge power output, respectively, and $c$ is the marginal discharge cost. The charge/discharge power levels and the final storage SoC are constrained by a feasibility set $\mathcal{E}(e_{t-1})$, which depends on the initial SoC, $e_{t-1}$, at the beginning of time period $t$. The feasibility set $\mathcal{E}(e_{t-1})$ includes the following constraints:
\begin{gather}
    0 \leq b_t \leq P\text{,}\; 0\leq p_t \leq P \label{p0_c1} \\
    \text{$p_t = 0$ if $\lambda_t < 0$} \label{p0_c2}\\
    e_t - e_{t-1} = -p_t/\eta + b_t\eta \label{p0_c3}\\
    0 \leq e_t \leq E \label{p0_c4}
\end{gather}
\end{subequations}\label{p0}
where \eqref{p0_c1} constraints on the storage charge and discharge power output. \eqref{p0_c2} is a relaxed form of the constraint that enforces the energy storage not charging and discharging simultaneously. Negative price is the necessary condition for storage to charge and discharge simultaneously in price arbitrage~\cite{xu2020operational}. \eqref{p0_c3} models the SoC evolution constraint with efficiency $\eta$ and \eqref{p0_c4} constraints the storage SoC. 

We define marginal opportunity cost function $q_t$ as first order derivative of the optimal opportunity cost:
\begin{align}
    q_{t}(e) &= \frac{\partial}{\partial e}Q_t(e)
\end{align}
In a deterministic setting with no price uncertainty, often referred to as a hindsight scenario, we can calculate $q_t$ recursively as a closed form~\cite{zheng2022arbitraging}:
\begin{align}\label{eq:vf}
    &q_{t-1}(e) = \nonumber\\
    &\begin{cases}
    q_{t}(e+P\eta)  & \text{if $\lambda_{t}\leq q_{t}(e+P\eta)\eta$} \\
    \lambda_{t}/\eta  & \text{if $ q_{t}(e+P\eta)\eta < \lambda_{t} \leq q_{t}(e)\eta$} \\
    q_{t}(e) & \text{if $ q_{t}(e)\eta < \lambda_{t} \leq [q_{t}(e)/\eta + c]^+$} \\
    (\lambda_{t}-c)\eta & \text{if $ [q_{t}(e)/\eta + c]^+ < \lambda_{t}$} \\
    & \quad\text{$ \leq [q_{t}(e-P/\eta)/\eta + c]^+$} \\
    q_{t}(e-P/\eta) & \text{if $\lambda_{t} > [q_{t}(e-P/\eta)/\eta + c]^+$} 
    \end{cases}
\end{align}

Using first order condition, we can design hindsight optimal bids for charging and discharging base on marginal value counting in both physical and opportunity costs:
\begin{subequations}
\begin{align}
    \partial(cp_t - Q_t(e_t))/{\partial p_{t}} &=  {\eta}q_{t}(e_t)/\eta \\
    \partial(cp_t - Q_t(e_t))/{\partial b_{t}} &=  -\eta q_{t}(e_t)
\end{align}\label{eq:bidcurve}
\end{subequations}

\noindent where~\eqref{eq:bidcurve} represents the hindsight optimal bid curve as a function of SoC.
\vspace{0.2cm}
\section{Analysis Results}\label{results}


\subsection{Availability withholding}

\begin{figure}[htbp]
\setlength{\abovecaptionskip}{-0.1cm}  
    \setlength{\belowcaptionskip}{-0.1cm} 
    \centering
\includegraphics[width=\columnwidth]{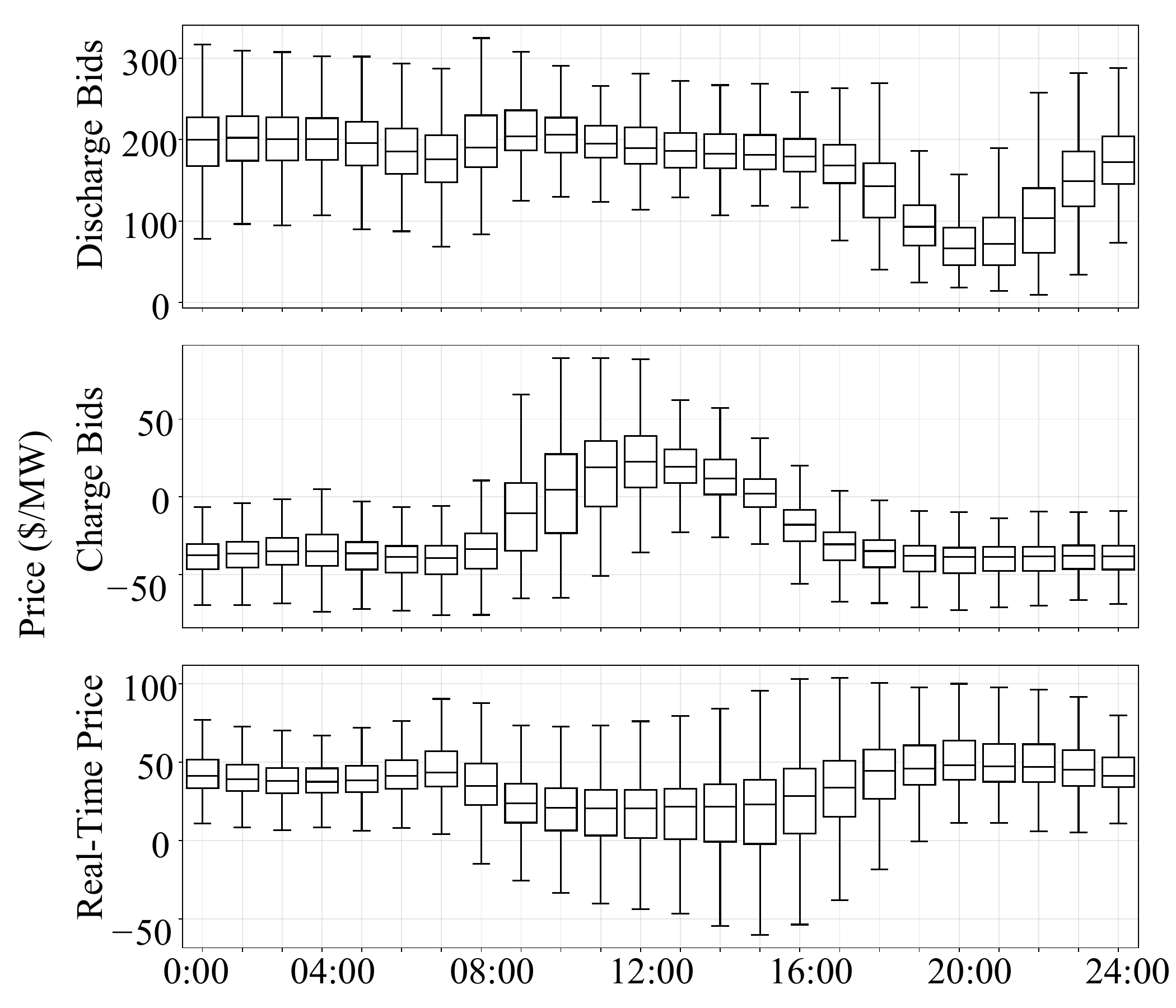}
    \caption{Average hour-of-day weighted RTPD discharge (top) and charge (middle) bid distributions and the RTPD price (bottom).}
    \label{HourofDayBox}
\end{figure}

Fig.~\ref{HourofDayBox} displays hour-of-day bids aggregated over the entire dataset, where we see a strong temporal dependence for both charge and discharge bids in RTPD. The discharge bids usually held around \$200/MW, except for slight reductions during early mornings and major reductions during the late evening, while charge bids held around \$-50/MW but increased during noon. Both trends negatively correlate with the duck-curve RTPD price pattern with price hikes during early morning and late evening, and price valleys during the noon due to high solar generation. Notably, the late evening has a significantly higher occurrence of extreme price spikes, which are not visualized in Fig.~\ref{HourofDayBox}. These bidding patterns represent a clear intention of availability withholding that storage strategically charges during low-price noon periods and discharges during late evening while setting bids to extremely high (low for charge bids) to withhold from market clearing.


\begin{figure}[htbp]
\setlength{\abovecaptionskip}{-0.1cm}  
    \setlength{\belowcaptionskip}{-0.1cm} 
    \centering
\includegraphics[width=\columnwidth]{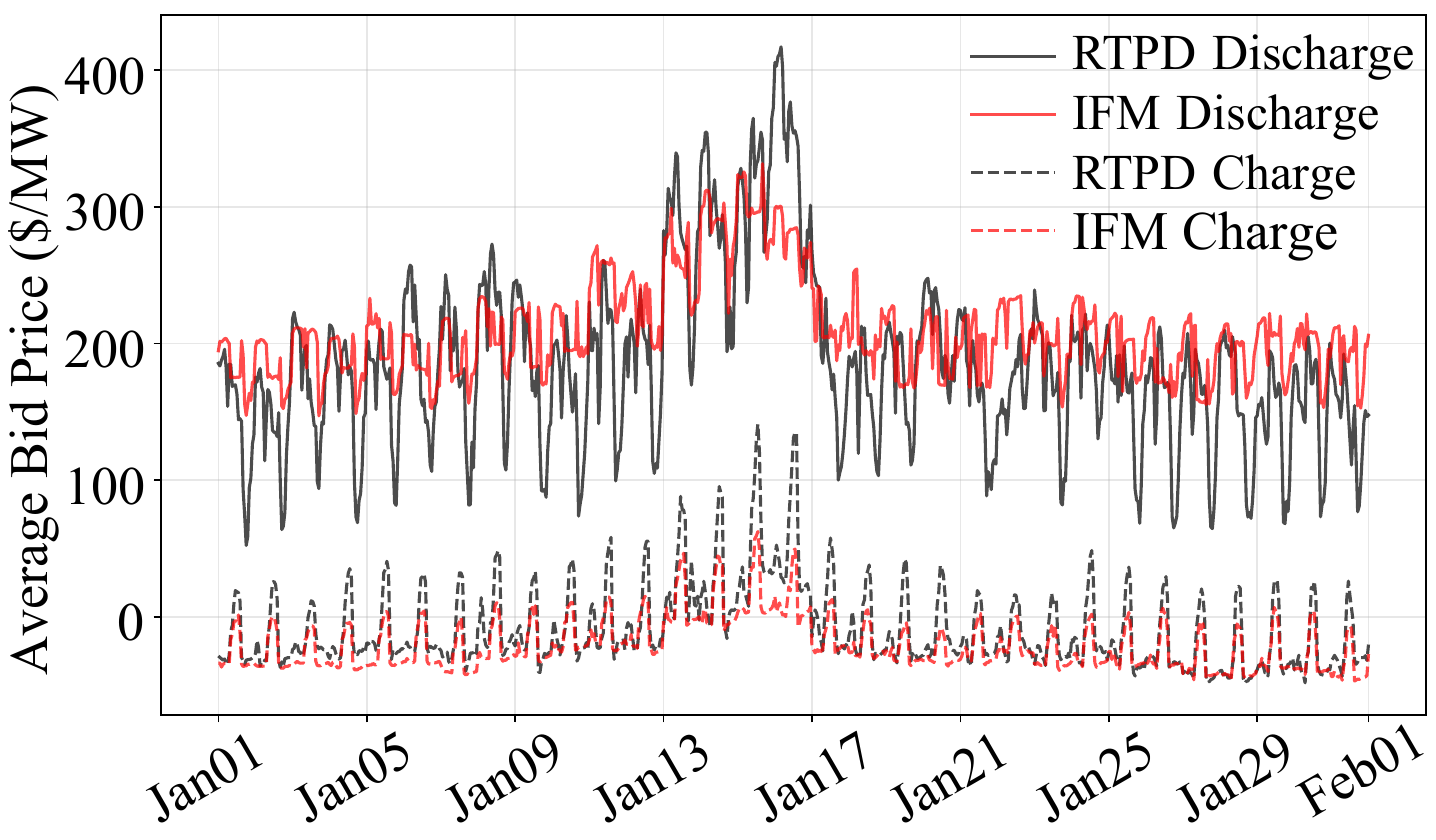}
    \caption{Weighted average hourly charge and discharge bids in IFM and RTPD during January 2024}
    \label{BidsJan}
\end{figure}

Fig.~\ref{BidsJan} shows the weighted average charge and discharge bids in IFM and RTPD. We observe IFM discharge bids at higher values than RTPD ones during $79$\% of the dataset, and IFM charge bids at lower values than RTPD ones for $87$\% of the dataset. In particular, IFM discharge bid prices were not reduced in the late evening, and the charge bids were not raised high enough during midday. This shows BESS operators are withholding capacity in the day-ahead market to more effectively leverage fluctuations in real-time markets. 

\subsection{Historical and Optimal Bid Comparison}

\begin{figure}[htbp]
\setlength{\abovecaptionskip}{-0.1cm}  
    \setlength{\belowcaptionskip}{-0.1cm} 
    \centering
\includegraphics[ width=\columnwidth]
{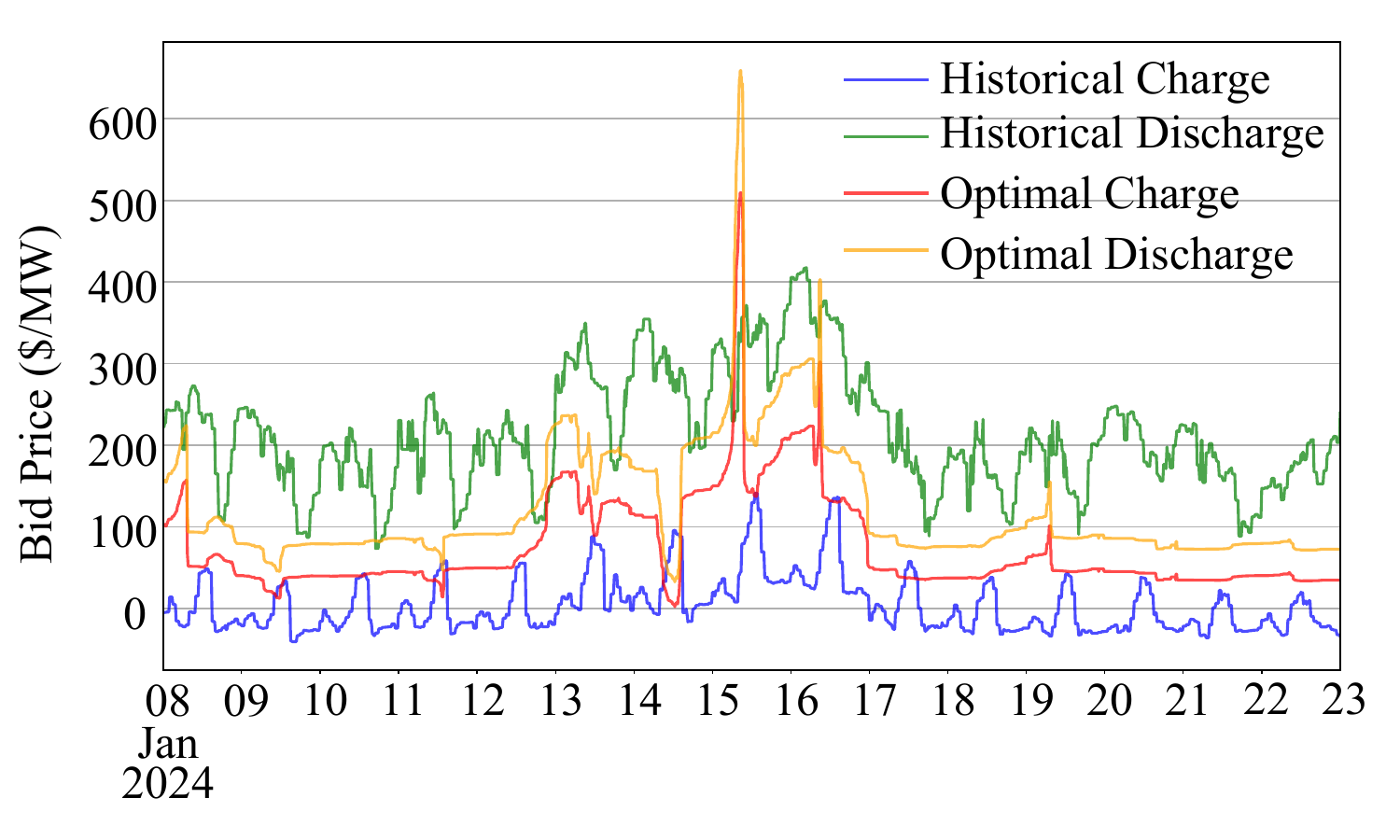}
    \caption{Hourly optimal hindsight and historical weighted average bids during January 2024}
    \label{OptimalHistoricalBids}
\end{figure}

We analyze historical bids by comparing them to optimally generated energy storage bids in hindsight. Figure~\ref{OptimalHistoricalBids} illustrates the optimal charge and discharge bids over a two-week period in January 2024. The historical charge and discharge bids exhibit a consistent daily pattern, whereas the optimal bids appear noticeably smoother. On most days, the optimal hindsight bids align closely with the lower and upper bounds of the average historical discharge and charge bids, respectively. This suggests that BESS operators selectively bid capacity, with charging concentrated in the middle of the day and discharging primarily in the evening. Notably, the lower limit of daily discharge bids (typically occurring in the late evening) and the higher limit of charge bids (mid-day) closely approximate the optimal discharge and charge bids. This alignment indicates that storage participants have a sound understanding of competitive economic bid values necessary for dispatch. Additionally, the high bid values outside these periods appear strategically designed to withhold capacity.


\begin{figure}[htbp]
\setlength{\abovecaptionskip}{-0.1cm}  
    \setlength{\belowcaptionskip}{-0.1cm} 
    \centering
\includegraphics[width=1.02\columnwidth]{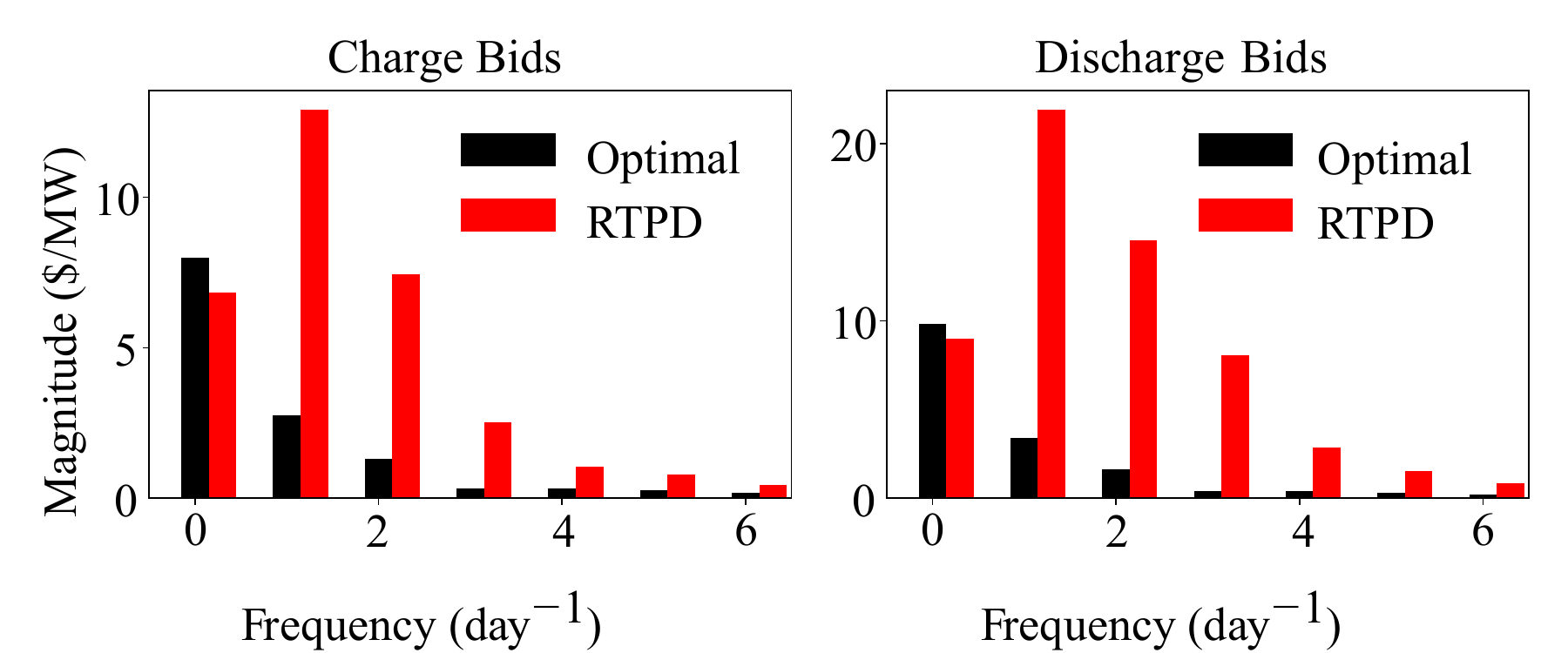}
    \caption{Spectral decomposition of hourly optimal hindsight and historical weighted average RTPD charge (left) and discharge (right) bids}
    \label{OptimalHistoricalFFT}
\end{figure}

This periodic trend can be further highlighted through a fast Fourier transform of real-time historical bids and optimal hindsight bids as in Fig.~\ref{OptimalHistoricalFFT}. Here, the height of each bar indicates the approximate magnitude of the dominant frequency within the respective bin. From this spectral decomposition, there is an extremely strong daily frequency signal supporting the daily cycling of historical bids that are exhibited in neither optimal charge nor discharge bids.

The sum of the dominant frequencies approximately represents the amplitude of oscillations from the decomposed signal. For historical bids, these magnitudes sum to approximately $30$ and $60$ \$/MW. This indicates that BESS operators are withholding significant discharge and charge capacity by, on average, varying bids by nearly $120$ and $60$ \$/MW daily, respectively. These results corroborate the variations in average charge and discharge bids observed in Fig.~\ref{BidsJan}.

\subsection{Price Spike Days}

\begin{figure}[htbp]
\setlength{\abovecaptionskip}{-0.1cm}  
    \setlength{\belowcaptionskip}{-0.1cm} 
    \centering
\includegraphics[trim = 0mm 5mm 0mm 5mm, clip, width=\columnwidth]{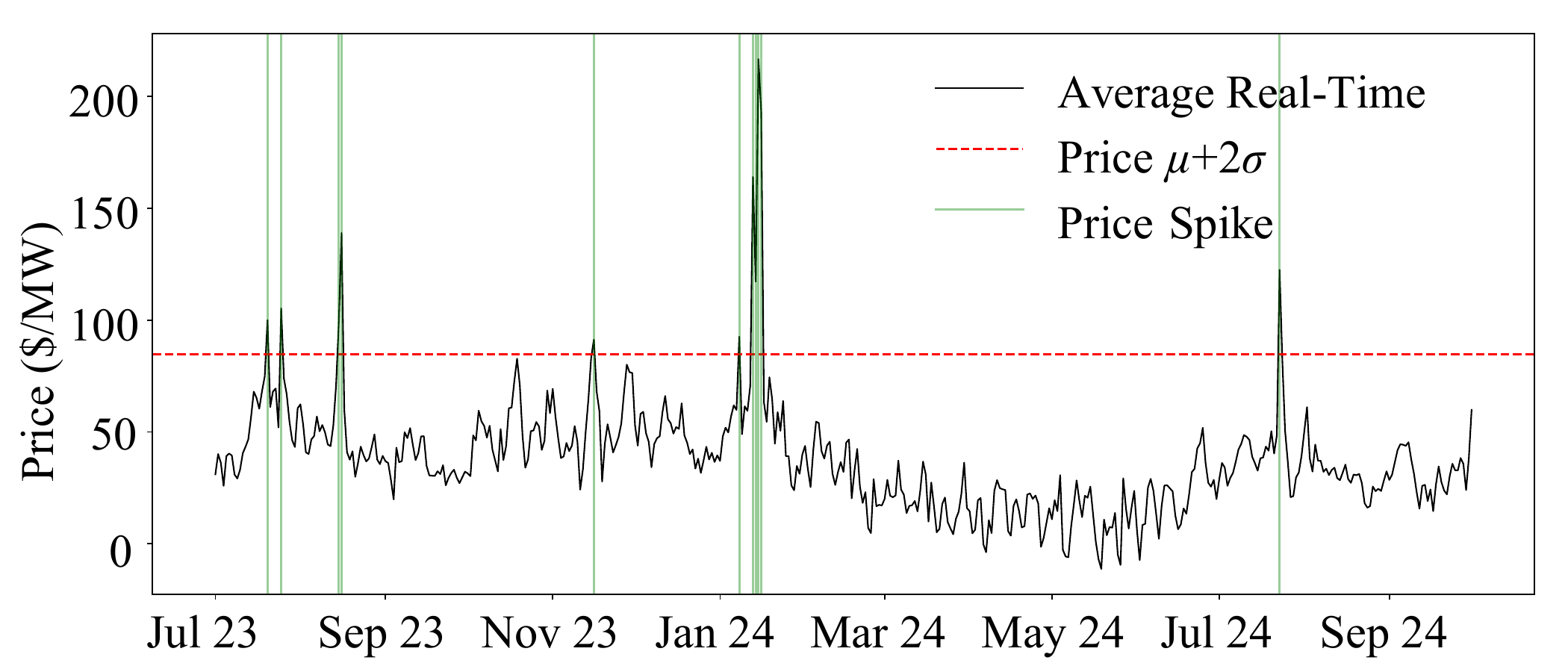}
    \caption{Average RTPD prices across CAISO with identified price spikes}
    \label{PriceSpikes}
\end{figure}

We study the bids during high-price periods to better understand the ability of BESS operators to predict price spikes and exercise market power. We identify price spike days as days where the price of electricity during that day exceeds two standard deviations above the mean daily price over the entire dataset. Price spike days are identified for all three price aggregations--- instantaneous minimum, maximum, and average--- in both the real-time and day-ahead markets. Fig.~\ref{PriceSpikes} shows the average RTPD price and eleven price spike dates identified over this period, in which bid spread price spikes are the ones that exceed two standard deviations above the mean. 

\begin{figure}[htbp]
\setlength{\abovecaptionskip}{-0.1cm}  
    \setlength{\belowcaptionskip}{-0.1cm} 
    \centering
\includegraphics[width=\columnwidth]
{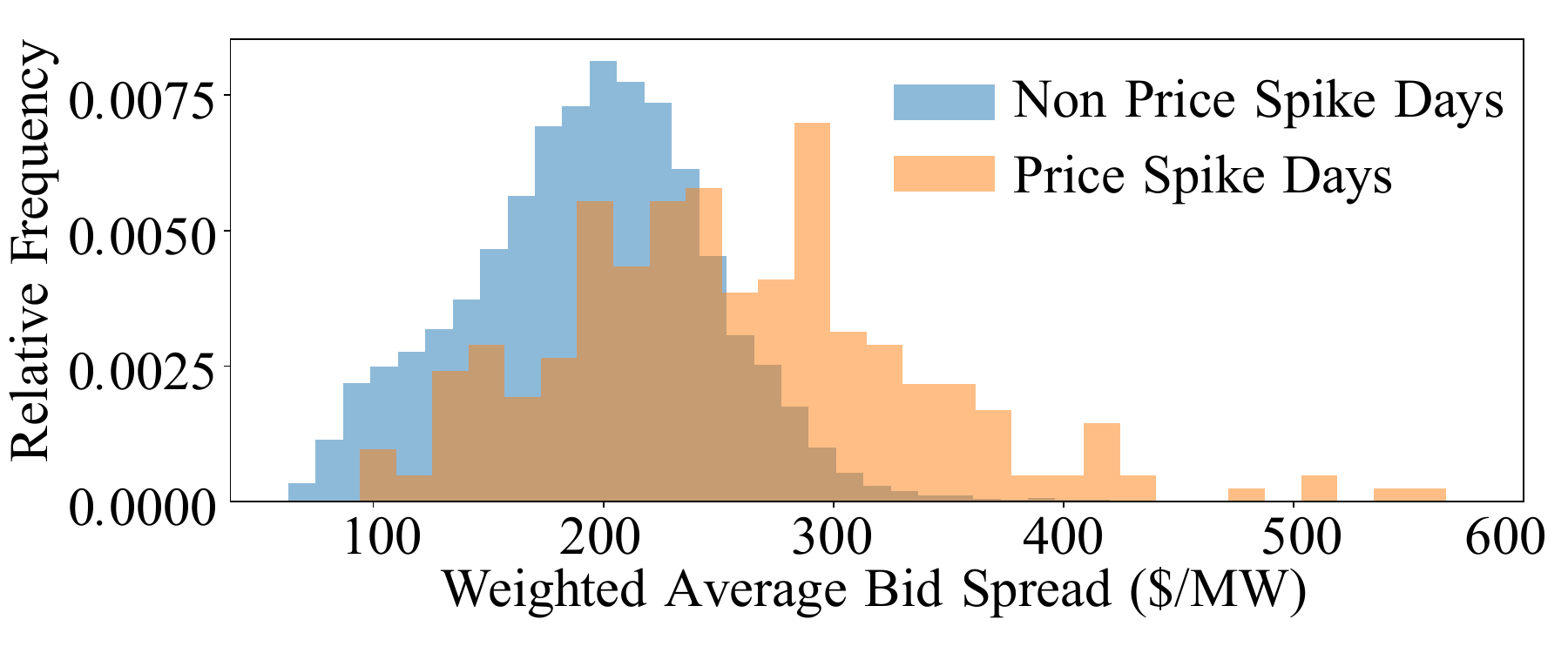}
    \caption{Hourly RTPD bid spread distribution for price spike and non-price spike days}
    \label{PriceSpikesDists}
\end{figure}

From the defined price spike days, we can form distributions of hourly bid spreads both during and excluding price spike days. In Fig.~\ref{PriceSpikesDists} a clear shift in the distribution of bid spreads is visible between price spike days, identified through average real-time prices, and RTPD bid spikes.  We test the likelihood of these distributions being sampled from the same population using the Kolmogorov-Smirnov (K-S) test~\cite{KSTest}, a distribution-free non-parametric statistical test. For price spreads in IFM and RTPD, the K-S test p-values are all less than $10^{-30}$ showing significant differences in distributions.


For all price spike identification methods and corresponding bid spreads, we find hourly bid spreads vary significantly ($p < 10^{-30}$) between price spike and non price spike days. This difference suggests a strong correlation between more extreme economic withholding of capacity and price spike days. As expressed previously, BESS does not face the risks of high costs while withholding capacity, so it is advantageous for it to withhold capacity more frequently in an attempt to capture high profits during price spikes. While the direct effects of this economic withholding are difficult to study without more granular data, we can inspect an identified price spike day to illustrate issues that may arise from such extreme withholding. 

\begin{figure}[htbp]
\setlength{\abovecaptionskip}{-0.1cm}  
    \setlength{\belowcaptionskip}{-0.1cm} 
    \centering
\includegraphics[trim = 0mm 5mm 0mm 5mm, clip, width=\columnwidth]{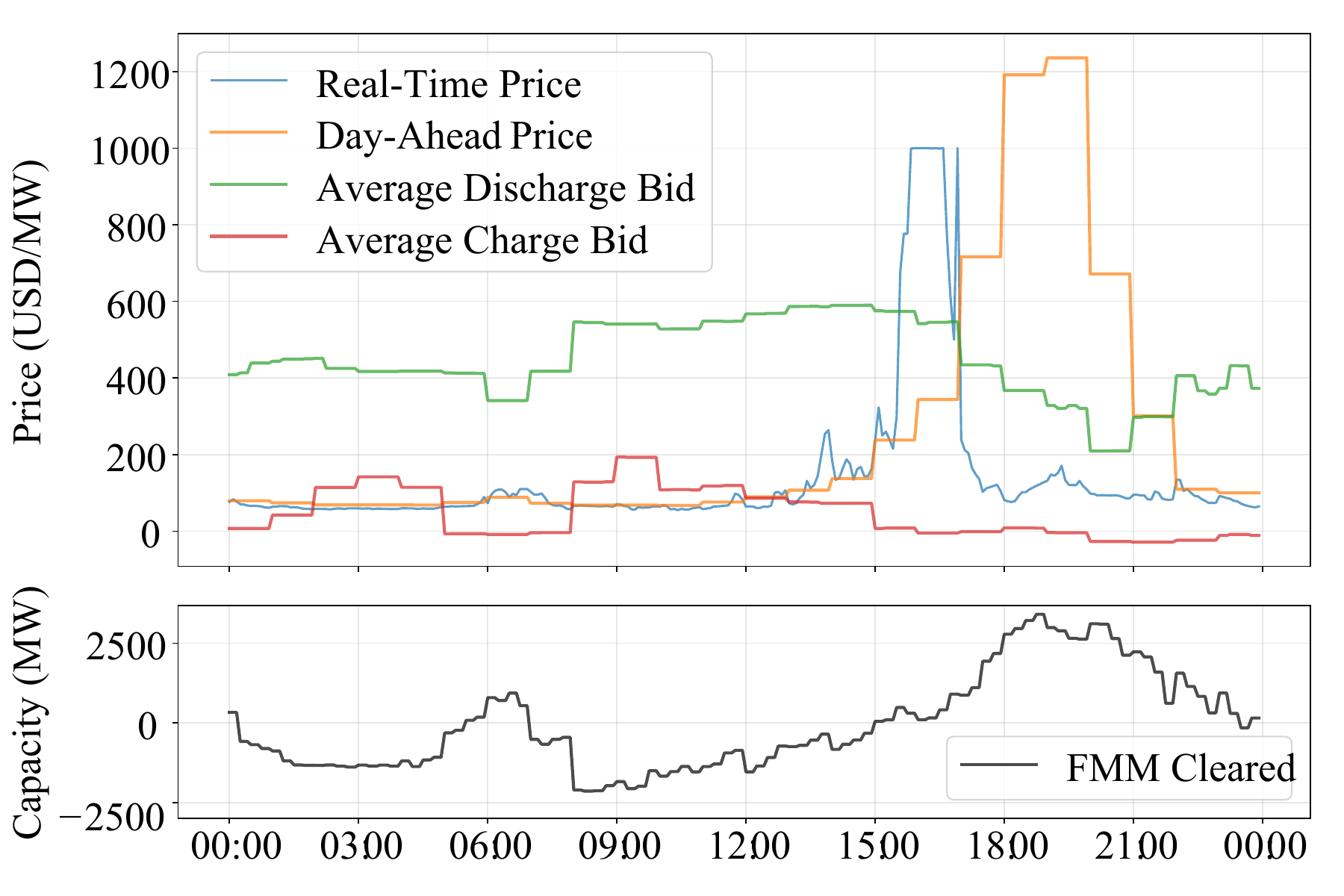}
\caption{Average Real-Time and Day-Ahead prices along with average weighted RTPD charge and discharge bids (top) and the storage capacity cleared in real-time markets (bottom) on August 16, 2023.}
\label{ExamplePriceSpikeDay}
\end{figure}

\subsection{Evidence of Inefficient Withholding}

Fig.~\ref{ExamplePriceSpikeDay} shows one of the price spike days where a spike in the real-time price without energy storage response. During this system-wide price spike, at about 4-5~PM, there was near no response reflected in average discharge bids barely decreasing from approximately 575 \$/MW to 550 \$/MW. This stability of inflated bids is observed despite energy storage units as a whole charging between 7~AM and 3~PM, indicating that even with storage capacity available for discharge during the price spike, storage operators likely chose to withhold capacity in anticipation of higher future prices.

The cleared storage capacity shown in Fig.~\ref{ExamplePriceSpikeDay} confirms our analysis above. On average, less than 500~MW of storage was discharged during the price spike while subsequently, between 6 and 9~PM, over 2500~MW of capacity was discharged on average despite prices falling to around 100 \$/MW. Such rigid storage bid patterns likely lead to inefficiencies in energy storage dispatch and elevated system operating costs.



\vspace{0.2cm}
\section{Discussion and Conclusion}\label{conclusion}

We conduct an in-depth analysis of historical energy storage bids in CAISO from July 1, 2023, to October 1, 2024. Our findings reveal daily bidding patterns that suggest economic withholding, a strategy often used by storage participants since they are prohibited from physical withholding. Specifically, in day-ahead markets, storage bids frequently withhold capacity to strategically participate in the more volatile real-time markets for greater profits. In real-time markets, storage bids reflect selective availability, typically charging during mid-day and discharging in the late evening when price spikes are most probable. Our analysis also shows that storage participants are highly responsive to price spike days, adjusting discharge bid values in response to system price surges. Overall, we conclude that the primary motivation behind current storage bid strategies in California is to withhold charge and discharge availability, rather than to reflect accurate price predictions or future opportunity value modeling.

While this strategy may align with current system conditions and storage penetration levels, there are several potential adverse impacts. First, the high withholding bid values limit storage availability during periods when storage operators are not interested in charging or discharging. For example, on August 16, an early-evening price spike saw limited storage response due to high discharge bids, likely because storage operators did not anticipate this spike. Although this withholding may not directly threaten system reliability (as storage is still physically available), it can drive up system costs by withholding storage from potential discharge opportunities outside the late evening.

Second, withholding availability from day-ahead markets may increase system carbon emissions, as unit commitment may optimize to commit more conventional generators \cite{qin2023role}. This misalignment of incentives deserves attention: current markets provide no price signal tied to emissions, prompting storage participants to avoid the day-ahead market to capitalize on real-time price volatility. This underscores the need for either incentives to encourage day-ahead market participation for storage or regulations to limit storage withholding patterns.

Third, we emphasize the difficulty in distinguishing between market power and legitimate withholding behaviors in energy storage. Many CAISO participants withhold their availability based on timing preferences for charge or discharge, without a clear physical or economic benchmark for bids. This makes identifying market power abuse challenging, unlike conventional generators, where regulators can use fuel prices as a baseline. Given that storage participants have similar access to market power practices as other market actors, these findings highlight the need for regulatory updates to address potential market power in storage.

As BESS capacity in CAISO continues to expand, market regulators need to establish effective practices to identify and monitor storage market power. Recent studies provide insights into this area. For instance, while BESS operators must account for uncertainties in bid design, making it unfair to directly compare with hindsight optimal bids~\cite{XuHobbs2024}, one study suggests that storage discharge bids should generally not exceed expected daily price spikes~\cite{zhou2024energy}, indicating that dramatic bid inflation outside late evening peak periods may be unwarranted. Another study establishes conditions for detecting market power through ex-post analysis of storage cycling patterns~\cite{wu2024}. 
Together, these studies outline pathways for developing new regulatory frameworks for storage market power.

\bibliographystyle{IEEEtran}
\bibliography{IEEEabrv,main}

\begin{thebibliography}{10}
\providecommand{\url}[1]{#1}
\csname url@samestyle\endcsname
\providecommand{\newblock}{\relax}
\providecommand{\bibinfo}[2]{#2}
\providecommand{\BIBentrySTDinterwordspacing}{\spaceskip=0pt\relax}
\providecommand{\BIBentryALTinterwordstretchfactor}{4}
\providecommand{\BIBentryALTinterwordspacing}{\spaceskip=\fontdimen2\font plus
\BIBentryALTinterwordstretchfactor\fontdimen3\font minus \fontdimen4\font\relax}
\providecommand{\BIBforeignlanguage}[2]{{%
\expandafter\ifx\csname l@#1\endcsname\relax
\typeout{** WARNING: IEEEtran.bst: No hyphenation pattern has been}%
\typeout{** loaded for the language `#1'. Using the pattern for}%
\typeout{** the default language instead.}%
\else
\language=\csname l@#1\endcsname
\fi
#2}}
\providecommand{\BIBdecl}{\relax}
\BIBdecl

\bibitem{CAISO_ES_2023}
\BIBentryALTinterwordspacing
{California ISO}, ``2023 special report on battery storage,'' Jul. 2024. [Online]. Available: \url{https://www.caiso.com/documents/2023-special-report-on-battery-storage-jul-16-2024.pdf}
\BIBentrySTDinterwordspacing

\bibitem{zheng2023energy}
N.~Zheng, X.~Qin, D.~Wu, G.~Murtaugh, and B.~Xu, ``Energy storage state-of-charge market model,'' \emph{IEEE Transactions on Energy Markets, Policy and Regulation}, vol.~1, no.~1, pp. 11--22, 2023.

\bibitem{zhou2024energy}
Z.~Zhou, N.~Zheng, R.~Zhang, and B.~Xu, ``Energy storage market power withholding bounds in real-time markets,'' in \emph{Proceedings of the 15th ACM International Conference on Future and Sustainable Energy Systems}, 2024, pp. 215--225.

\bibitem{harvey2001market}
S.~M. Harvey and W.~W. Hogan, ``Market power and withholding,'' \emph{Harvard Univ., Cambridge, MA}, 2001.

\bibitem{caiso_phase4_2023}
\BIBentryALTinterwordspacing
{California Independent System Operator}, ``Final proposal: Energy storage and distributed energy resources phase 4 - default energy bid,'' 2023, accessed: 2024-11-09. [Online]. Available: \url{https://stakeholdercenter.caiso.com/initiativedocuments/finalproposal-energystorage-distributedenergyresourcesphase4-defaultenergybid.pdf}
\BIBentrySTDinterwordspacing

\bibitem{sioshansi_when_2014}
R.~Sioshansi, ``When energy storage reduces social welfare,'' \emph{Energy Economics}, vol.~41, pp. 106--116, Jan. 2014.

\bibitem{williams_electricity_2022}
O.~Williams and R.~Green, ``Electricity storage and market power,'' \emph{Energy Policy}, vol. 164, p. 112872, May 2022.

\bibitem{mohsenian-rad_coordinated_2016}
H.~Mohsenian-Rad, ``Coordinated {Price}-{Maker} {Operation} of {Large} {Energy} {Storage} {Units} in {Nodal} {Energy} {Markets},'' \emph{IEEE Transactions on Power Systems}, vol.~31, no.~1, pp. 786--797, Jan. 2016.

\bibitem{ye_investigating_2019}
Y.~Ye, D.~Papadaskalopoulos, R.~Moreira, and G.~Strbac, ``Investigating the impacts of price‐taking and price‐making energy storage in electricity markets through an equilibrium programming model,'' \emph{IET Generation, Transmission \& Distribution}, vol.~13, no.~2, pp. 305--315, Jan. 2019.

\bibitem{chabok_assessment_2020}
H.~Chabok, M.~Roustaei, M.~Sheikh, and A.~Kavousi-Fard, ``On the assessment of the impact of a price-maker energy storage unit on the operation of power system: {The} {ISO} point of view,'' \emph{Energy}, vol. 190, p. 116224, Jan. 2020.

\bibitem{bhattacharjee_energy_2022}
S.~Bhattacharjee, R.~Sioshansi, and H.~Zareipour, ``Energy {Storage} {Participation} in {Wholesale} {Markets}: {The} {Impact} of {State}-of-{Energy} {Management},'' \emph{IEEE Open Access Journal of Power and Energy}, vol.~9, pp. 173--182, 2022.

\bibitem{caiso_daily_storage_reports}
\BIBentryALTinterwordspacing
{California ISO}, ``Daily energy storage reports,'' 2024. [Online]. Available: \url{https://www.caiso.com/library/daily-energy-storage-reports}
\BIBentrySTDinterwordspacing

\bibitem{zheng2022arbitraging}
N.~Zheng, J.~J. Jaworski, and B.~Xu, ``Arbitraging variable efficiency energy storage using analytical stochastic dynamic programming,'' \emph{IEEE Transactions on Power Systems}, 2022.

\bibitem{xu2020operational}
B.~Xu, M.~Korp{\aa}s, and A.~Botterud, ``Operational valuation of energy storage under multi-stage price uncertainties,'' in \emph{2020 59th IEEE Conference on Decision and Control (CDC)}.\hskip 1em plus 0.5em minus 0.4em\relax IEEE, 2020, pp. 55--60.

\bibitem{KSTest}
\BIBentryALTinterwordspacing
D.~A. Darling, ``The kolmogorov-smirnov, cramér-von mises tests,'' \emph{The Annals of Mathematical Statistics}, vol.~28, no.~4, pp. 823--838, 1957. [Online]. Available: \url{http://www.jstor.org/stable/2237048}
\BIBentrySTDinterwordspacing

\bibitem{qin2023role}
X.~Qin, B.~Xu, I.~Lestas, Y.~Guo, and H.~Sun, ``The role of electricity market design for energy storage in cost-efficient decarbonization,'' \emph{Joule}, vol.~7, no.~6, pp. 1227--1240, 2023.

\bibitem{XuHobbs2024}
\BIBentryALTinterwordspacing
B.~Xu and B.~F. Hobbs, ``On truthful pricing of battery energy storage resources in electricity spot markets,'' \emph{Oxford Energy Forum}, no. 140, pp. 34--37, April 2024. [Online]. Available: \url{https://www.oxfordenergy.org/wpcms/wp-content/uploads/2024/04/OEF-140-Powering-the-Future.pdf}
\BIBentrySTDinterwordspacing

\bibitem{wu2024}
\BIBentryALTinterwordspacing
Y.~Wu, B.~Xu, and J.~Anderson, ``Market power and withholding behavior of energy storage units,'' in \emph{2024 63rd IEEE Conference on Decision and Control (CDC) (to appear)}, 2024. [Online]. Available: \url{https://arxiv.org/abs/2405.01442}
\BIBentrySTDinterwordspacing

\end{thebibliography}

\end{document}